\newcommand{\openone}{\leavevmode\hbox{\small1\normalsize\kern-.33em1}}
\newcommand{\Tr}{\mathop{\mathrm{Tr}}\nolimits}

\documentclass{elsart}
\usepackage{graphicx,amsfonts,amssymb,bm,subeqnarray}

\begin{document}

\begin{frontmatter}

\title{Quantum phase-space description of
light polarization}

\author{A. B. Klimov{}$^{\mathrm{a}}$,
J. Delgado{}$^{\mathrm{b}}$,
L. L. S\'anchez-Soto{}$^{\mathrm{b}}$}

\address{{}$^{\mathrm{a}}$
Departamento de F\'{\i}sica,
Universidad de Guadalajara,
Revoluci\'on~1500, 44420~Guadalajara,
Jalisco, Mexico}

\address{{}$^{\mathrm{b}}$
Departamento de \'Optica,
Facultad de F\'{\i}sica,
Universidad Complutense,
28040~Madrid, Spain}


\maketitle

\begin{abstract}
We present a method to characterize the
polarization state of a light field in the
continuous-variable regime. Instead of using
the abstract formalism of SU(2) quasidistributions,
we model polarization in the classical spirit
by superposing two harmonic oscillators of the
same angular frequency along two orthogonal axes.
By describing each oscillator by a $s$-parametrized
quasidistribution, we derive in a consistent way
the final function for the polarization. We
compare with previous approaches and discuss
how this formalism works in some relevant examples.
\end{abstract}

\end{frontmatter}

\section{Introduction}

Polarization is a fundamental property of
light, both in the quantum and in the
classical domain. In quantum optics,
polarization has been mainly examined in
the single-photon regime~\cite{Per93,Kwi95,Mat96,Mul97,Zei99,Tri01,Bar03}.
Nevertheless, different schemes have
been proposed~\cite{Alo98,Kor02} and
experimentally implemented~\cite{Gra87,Bow02}
to characterize the continuous-variable
limit of the quantum Stokes parameters.
We stress that these continuous-variable
polarization states can be carried by a bright
laser beam providing high bandwidth capabilities
and therefore faster signal transfer rates than
single photon systems. In addition, they retain
the single-photon advantage of not requiring the
universal local oscillator necessary for other
proposed continuous variable quantum networks.

Since the seminal paper of Wigner~\cite{Wig32},
and the major contributions of Moyal~\cite{Moy49},
Stratonovich~\cite{Str56}, and Berezin~\cite{Ber75},
it seems indisputable that phase-space methods,
based on using quasidistributions that reflect the
noncommutatibility of quantum observables,
constitute a valuable tool in examining continuous
variables in quantum optics~\cite{Hil84,Per98,Sch01}.

In particular, these methods have had great
success in analyzing one-mode fields; i. e.,
Heisenberg-Weyl quasidistributions representing
the quantum dynamics in the flat $q$-$p$  (or,
equivalently, $a$-$a^\ast$) space. Although not
so popular in the quantum-optics community, spinlike
systems, with the sphere $\mathcal{S}_2$ as phase
space, have been also discussed at length in this
framework~\cite{Aga81,Coh86,Var89,Wol96,Ram96,Aga97,Ata98,Bri99,Ben99}.
The resulting functions, naturally related to
the SU(2) dynamical group, have been used to
visualize, e. g., nonclassical properties of a
collection of two-level atoms~\cite{Dow94}.

Since the Stokes operators can be formally
identified with an angular momentum~\cite{Col70,Jau76,Chi93,Lui00},
one may naively expect a direct translation of
these SU(2) quasidistributions to the problem
of polarization. However, this is not the
case, mainly because they act on different
types of Hilbert space~\cite{Kar93}. We can
then conclude that the problem of an adequate
quasiclassical description of polarization of
light is still an open question~\cite{Kar92,Lui05}.

The Stokes operators are a particular case of
the Schwinger map~\cite{Sch60}, with two
kinematically independent oscillators.
In this spirit, it has been recently shown
that the Stokes operators are the constants
of motion of the two-dimensional isotropic
harmonic oscillator~\cite{Mot04}. This reflects
the fact that the polarization of a classical
field can be adequately viewed as the Lissajous
figure traced out by the end of the electric
vector of a monochromatic field~\cite{Bro98}.
In sharp contrast, in quantum optics the
probability distribution for the electric field
can be very far from having an elliptical
form~\cite{Pol95,Lui02}.

We wish to investigate this point from the
perspective of quasidistributions.  We show that
one can start from the $s$-ordered quasidistributions
from two kinematically independent oscillators: by
eliminating an unessential common phase, we
get well-behaved quasidistributions on the
Poincar\'e sphere. We apply the resulting family
of polarization quasidistributions to some
relevant states, and conclude that they constitute
an appropriate tool to deal with such a basic variable.

\section{Phase-space representation of a
harmonic oscillator}

To keep the discussion as self-contained
as possible, we first briefly summarize
the essential ingredients of phase-space
functions for a harmonic oscillator that
we shall need for our purposes.

In the Hilbert space $\mathcal{H}$, the state
of the system  is fully represented by its
density operator $\hat{\varrho}$. In the
phase-space formalism, $\hat{\varrho}$
is mapped by a family of functions (quasidistributions)
$W^{(s)} (\alpha)$ onto the classical phase
space $X$ ($\alpha \in X$). This map is usually
implemented by the generalized Weyl rule~\cite{Bri99}
\begin{equation}
W^{(s)} (\alpha)= \Tr  [ \hat{\varrho} \,
\hat{w}^{(s)} (\alpha) ] ,
\end{equation}
where the generating kernel $\hat{w}^{(s)}
(\alpha)$ fulfill the general properties
\begin{equation}
\begin{array}{l}
\hat{w}^{(s)} (\alpha) =
[ \hat{w}^{(s)} (\alpha) ]^\dagger , \\
 \\
\displaystyle
\int_X d\mu (\alpha ) \, \hat{w}^{(s)}
(\alpha) = \hat{\openone} .
\end{array}
\end{equation}
The index $s$ that labels functions in the
family is related to the $s$ ordering.
The values $-1$, $+1$, and $0$ correspond
to the normal, antinormal, and symmetric
ordering, respectively, or equivalently to
the $P$, $Q$, and $W$ functions. We stress
that these quasidistributions can be determined
in practice by using simple and efficient
experimental procedures~\cite{Vog89,Smi93,Leo97,Wel99,Lvo01,Serge02}.
Moreover, they provide a simple measure
of the nonclassical behavior of quantum
states~\cite{Luk95,Lim96,Dod02,VWW01}.

Let us now turn to the outstanding case of a
harmonic oscillator represented by annihilation
and creation operators $\hat{a}$ and
$\hat{a}^\dagger$, that obey the canonical
commutation relation
\begin{equation}
\label{ccr}
[\hat{a}, \hat{a}^\dagger ] = \hat{\openone} .
\end{equation}
The phase space is the complex plane $\mathbb{C}$
and the invariant measure is $d\mu(\alpha ) =
d^2 \alpha/\pi^2$. The operator
\begin{equation}
\hat{D} (\alpha) = \exp(\alpha \hat{a}^\dagger
- \alpha^\ast \hat{a} )
\end{equation}
is the standard displacement operator in
the complex plane $\alpha$ and leads to
introduce the standard coherent states as
\begin{equation}
| \alpha \rangle = \hat{D} (\alpha)
| 0 \rangle ,
\end{equation}
where $| 0 \rangle$ denotes the ground
state. In this case, the kernel $\hat{w}^{(s)}
(\alpha)$ is exactly the Cahill-Glauber kernel
\begin{equation}
\label{wCG}
\hat{w}^{(s)} (\alpha) = \frac{1}{\pi^2}
\int_{\mathbb{C}} d^2 \beta \, e^{ - s |\beta|^2/2}
\,\exp(\alpha \beta^\ast - \alpha^\ast \beta)
\, \hat{D} (\alpha) .
\end{equation}
It is often useful to represent this operator in
a somewhat different form. After some calculations,
it turns out that (\ref{wCG}) may be rewritten
as~\cite{VWW01}
\begin{equation}
\label{w1}
\hat{w}^{(s)}(\alpha ) = \frac{2}{1-s}
\hat{D} (\alpha) \ \left ( \frac{s+1}{s-1}
\right )^{\hat{a}^\dagger \hat{a}} \
\hat{D}^\dagger (\alpha ) ,
\end{equation}
which can also be represented in a disentangled form
\begin{equation}
\hat{w}^{(s)} (\alpha ) = \frac{2}{1-s} \,
e^{ - \frac{2|\alpha |^2}{1-s}}
\left ( \frac{s+1}{s-1} \right )^{\hat{a}^\dagger \hat{a}}
\exp \left ( -\frac{2 \alpha}{1+s} \hat{a}^\dagger \right)
\exp \left ( \frac{2\alpha^\ast}{1-s} \hat{a} \right ) .
\end{equation}
The reader is referred to e.g. Ref.~\cite{Per98} to
see how this $s$-parametrized representation works for
some elementary field states.

\section{Phase-space description of two
orthogonal harmonic oscillators}

In classical optics, the superposition of two
oscillations of the same angular frequency $\omega$,
one along the horizontal axis $H$ and the other along
the vertical axis $V$, results in an elliptical motion.
This is the simplest Lissajous figure and is the basic
physics behind the notion of light polarization.

To translate this picture into quantum optics, we
assume that the two  oscillators are represented by
the complex amplitude operators $\hat{a}_H$
($\hat{a}_H^\dagger$) and $\hat{a}_V$ ($\hat{a}_V^\dagger$),
fulfilling (\ref{ccr}); i. e.,
\begin{equation}
[\hat{a}_j, \hat{a}_k^\dagger] =
\delta_{jk} \hat{\openone} \, ,
\qquad \qquad
j, k = H, V .
\end{equation}
In phase space, these oscillators can be appropriately
described by the product of the corresponding kernel
operators
\begin{equation}
\label{prodw1}
\hat{w}^{(s)} ( \alpha_H, \alpha_V )  =
\hat{w}^{(s)} ( \alpha_H ) \,
\hat{w}^{(s)} ( \alpha_V ) .
\end{equation}
A Lissajous figure needs only three independent
quantities to be fully characterized: the amplitudes
of each oscillator and the relative phase between them.
We therefore introduce the parametrization
\begin{equation}
\alpha_H = r e^{i \zeta}
\cos (\theta/2)  ,
\qquad
\alpha_V = r e^{i \zeta} e^{- i \phi}
\sin (\theta /2)  ,
\end{equation}
where
\begin{equation}
r^2 = |\alpha_H |^2 + |\alpha_V |^2
\end{equation}
is a radial variable related with the global
intensity. The parameters $\theta$ and $\phi$
can be interpreted as the polar and azimuthal
angles, respectively, on the Poincar\'e sphere
$\mathcal{S}_2$. In terms of this parametrization,
equation~(\ref{prodw1}) can be recast as
\begin{eqnarray}
\hat{w}^{(s)}(\alpha_H, \alpha_V) & = &
\left ( \frac{2}{1-s} \right )^2
e^{ - \frac{2r^2}{1-s}}
\left ( \frac{s+1}{s-1} \right )^{\hat{N}}
\nonumber \\
& \times &
\hat{U}(\theta,\phi) \,
\left [
\exp \left ( -\frac{2r}{1+s} e^{i \zeta}
\hat{a}_H^\dagger \right )
\exp \left ( \frac{2r}{1-s} e^{-i\zeta}
\hat{a}_H \right )
\right ] \,
\hat{U}^\dagger (\theta,\phi),
\nonumber \\
\end{eqnarray}
where
\begin{equation}
\hat{N} = \hat{a}_H^\dagger \hat{a}_H +
\hat{a}_V^\dagger \hat{a}_V
\end{equation}
is the operator representing the total number of
excitations and
\begin{equation}
\hat{U} (\theta,\phi) =
\exp \left[ \frac{\theta}{2}
\left ( e^{-i\phi}
\hat{a}_H  \hat{a}_V^\dagger -
e^{i\phi} \hat{a}_V \hat{a}_H^\dagger
\right ) \right ] .
\end{equation}
We next proceed to integrate over the physically
irrelevant global phase $\zeta$\footnote{Note, in passing,
that this equation can be expressed compactly as
\begin{eqnarray}
\hat{w}^{(s)}(\theta, \phi, r) & = &
\left ( \frac{2}{1-s} \right )^2
e^{-\frac{2r^2}{1-s}}
\left ( \frac{s+1}{s-1} \right )^{\hat{N}}
\nonumber \\
& \times & \hat{U}(\theta,\phi) \ : J_{0}
\left ( 4 r \sqrt{\frac{\hat{a}_H^{\dagger} \hat{a}_H}
{1-s^2}} \right) :
\hat{U}^\dagger (\theta,\phi), \nonumber
\end{eqnarray}
where $: \ :$ means normal ordering and $J_0$ denotes the Bessel
function of first kind and zero order.}:
\begin{eqnarray}
\hat{w}^{(s)} (r, \theta, \phi) & = &
\frac{1}{2\pi} \int_{-\pi}^{\pi} d\zeta \
\hat{w}^{(s)}(\alpha_H, \alpha_V) =
\left( \frac{2}{1-s} \right )^2
e^{\frac{2 r^2}{1-s}}
\left ( \frac{s+1}{s-1} \right )^{\hat{N}}
\nonumber \\
& \times &
\hat{U}(\theta,\phi) \
\left [
\sum_{\ell=0}^{\infty}
\frac{(- 1 )^\ell}{\ell!^2} r^{2\ell}
\left( \frac{4}{1-s^2} \right)^\ell
\hat{a}_H^{\dagger \ell}
\hat{a}_H^\ell
\right ]
\
\hat{U}^{\dagger}(\theta, \phi) \, .
\end{eqnarray}

Finally, we integrate over the radial variable $r$
(with the weight $2 r^2$) to obtain the phase-space
kernel over the sphere $\mathcal{S}_2$
\begin{eqnarray}
\hat{w}^{(s)}(\theta,\phi) & = & 2 \int_0^\infty dr \
r^3 \, \hat{w}^{(s)}(\theta,\phi,r) =
\left ( \frac{s+1}{s-1} \right )^{\hat{N}}
\nonumber \\
& \times &
\hat{U}(\theta,\phi) \
\left [
\sum_{\ell=0}^{\infty} (-1)^\ell \,
\frac{\ell + 1}{\ell!}
\left ( \frac{2}{1+s} \right )^\ell
\hat{a}_H^{\dagger \ell}  \hat{a}_H^\ell
\right ] \
\hat{U}^\dagger (\theta,\phi) \, .
\end{eqnarray}
If we observe that
\begin{equation}
\sum_{\ell=0}^{\infty} \frac{z^\ell}{\ell!} \;
\hat{a}_H^{\dagger \ell} \hat{a}_H^\ell =
(z + 1)^{\hat{a}_H^\dagger \hat{a}_H} \, ,
\end{equation}
we easily obtain
\begin{eqnarray}
\label{cenres}
\hat{w}^{(s)}(\theta, \phi) & = &
\left( \frac{s+1}{s-1} \right )^{\hat{N}}
\ \hat{U}(\theta,\phi) \
\left [
\left ( \frac{s-1}{s+1} \right )^{\hat{a}_H^\dagger \hat{a}_H}
\frac{2 \hat{a}_H^\dagger \hat{a}_H + 1-s}{1-s}
\right ] \
\hat{U}^{\dagger}(\theta,\phi)  \nonumber \\
& = &
\hat{U}(\theta,\phi)
\
\left [
\left( \frac{s+1}{s-1}\right )^{\hat{a}_V^\dagger \hat{a}_V}
\frac{2 \hat{a}_H^\dagger \hat{a}_H + 1-s}{1-s}
\right ] \
\hat{U}^{\dagger}(\theta,\phi) \, .
\end{eqnarray}
This is our central and compact result, that we shall work
out in detail in the rest of the paper.

For the antisymmetric ordering $(s = - 1)$, which corresponds
to the $Q$-function, (\ref{cenres}) reduces to
\begin{equation}
\hat{w}^{(-1)} (\theta, \phi) =
\hat{U}(\theta,\phi) \
\left [
\lim_{s\rightarrow -1}
\left( \frac{s+1}{s-1} \right )^{\hat{a}_V^\dagger \hat{a}_V}
( \hat{a}_H^\dagger \hat{a}_H + 1 )
\right ] \
\hat{U}^{\dagger}(\theta, \phi) \, .
\end{equation}
To proceed further we notice that
\begin{equation}
\label{proj}
\lim_{s \rightarrow -1}
\left ( \frac{s+1}{s-1} \right)^{\hat{a}_V^\dagger \hat{a}_V} =
\sum_{N=0}^\infty | N, 0 \rangle \ \langle N, 0 |
\end{equation}
is precisely the projector on the subspace with zero
excitations in the vertical mode $V$. Henceforth, we
shall denote by
\begin{equation}
\label{defbas}
| N, k \rangle := | N - k \rangle_H \otimes | k \rangle_V
\end{equation}
a state with $N - k$  excitations in the horizontal mode $H$
and $k$ in the vertical mode $V$. In view of (\ref{proj})
we have
\begin{equation}
\hat{w}^{(-1)}(\theta,\phi)  =
\sum_{N=0}^{\infty} ( N + 1) | N, \theta, \phi \rangle
\langle  N, \theta,\phi| \, ,
\end{equation}
where $|N,\theta,\phi \rangle$ are the standard
SU(2) coherent states~\cite{Per86}
\begin{eqnarray}
| N, \theta, \phi \rangle & = & \hat{U} (\theta, \phi)
| N, 0 \rangle  \nonumber \\
& = & \sum_{k=0}^N
\left ( \begin{array}{c}
N \\
k
\end{array}
\right )^{1/2}
\left ( \sin \frac{\theta}{2} \right )^k
\left ( \cos \frac{\theta}{2} \right )^{N - k}  \
e^{-i k \phi} | N, k \rangle \, .
\end{eqnarray}
The $Q$ function generated by this kernel reads as
\begin{equation}
\label{Q1}
Q(\theta,\phi)= \Tr [ \hat{\varrho}
\hat{w}^{(-1)}(\theta,\phi) ] =
\sum_{N=0}^{\infty} (N + 1) \
Q(N,\theta,\phi) \, ,
\end{equation}
where
\begin{equation}
Q(N,\theta,\phi) =
\langle N, \theta, \phi|
\hat{\varrho} |N,\theta, \phi \rangle \, .
\end{equation}
The normalization condition
\begin{equation}
\frac{1}{4 \pi}\int d\Omega \ Q (\theta, \phi)=1 \, ,
\end{equation}
where $d\Omega = \sin \theta  d\theta  d\phi$ is the
differential of solid angle, is automatically fulfilled.

Equation (\ref{Q1}) is a well-known result, which may
be derived from a variety of methods. The point we wish
to stress is here that (\ref{Q1}) involves only diagonal
elements between states with the same number of excitations.
Because of the lack of the off-diagonal contributions of
the form $\langle N, \theta, \phi | \hat{\varrho} |
N^\prime, \theta, \phi \rangle$ with $N \neq N^\prime$,
the $Q$ function takes the form of an average over
the subspaces with definite total number of excitations.
The role of the sum in $N$ is to remove the total intensity
from the description of the state.

We next pass to the symmetrical order ($s=0$),
which corresponds to the Wigner function. From
(\ref{cenres}) we immediately get
\begin{equation}
\label{Wop}
\hat{w}^{(0)}(\theta ,\phi ) =
\hat{U}(\theta ,\phi ) \
\left [
(-1)^{\hat{a}_V^\dagger \hat{a}_V}
( 2 \hat{a}_H^\dagger \hat{a}_H + 1 )
\right ] \
\hat{U}^\dagger (\theta ,\phi ).
\end{equation}
To properly transform the operator in brackets, it
proves convenient to introduce the following operators
\begin{eqnarray}
\label{gensu2}
\hat{J}_x & = & \frac{1}{2}
(\hat{a}_H^\dagger \hat{a}_V +
\hat{a}_V^\dagger \hat{a}_H ) \, ,
\nonumber \\
\hat{J}_y & = & \frac{1}{2i}
(\hat{a}_H^\dagger \hat{a}_V -
\hat{a}_V^\dagger \hat{a}_H ) \, , \\
\hat{J}_z & =  & \frac{1}{2}
( \hat{a}_H^\dagger \hat{a}_H -
\hat{a}_V^\dagger \hat{a}_V )  \, ,
\nonumber
\end{eqnarray}
which constitute the Schwinger map for two
independent oscillators~\cite{Sch60} and
are the generators of the group SU(2). Since
$\hat{U}(\theta ,\phi )$ is an element of SU(2),
it can be represented as~\cite{Per86}
\begin{equation}
\hat{U}(\theta ,\phi ) = e^{-i \phi \hat{J}_{z}}
\ e^{-i\theta \hat{J}_{y}} \
e^{i\phi \hat{J}_{z}}.
\end{equation}
Next, we note that $\hat{a}_H^\dagger \hat{a}_H =
\hat{N}/2 - \hat{J}_z$ and that
\begin{equation}
\hat{U}(\theta ,\phi ) \ \hat{J}_{z} \
\hat{U}^\dagger (\theta ,\phi ) =
\hat{\mathbf{J}} \cdot \mathbf{n} \,,
\end{equation}
where $\hat{\mathbf{J}} = (\hat{J}_x,
\hat{J}_y, \hat{J}_z)$ and $\mathbf{n} =
(\sin \theta \cos \phi, \sin \theta \sin \phi,
\cos \theta )$ is a unit vector on the sphere
$\mathcal{S}_2$. Then, equation~(\ref{Wop})
simplifies to
\begin{equation}
\label{wop}
\hat{w}^{(0)}(\theta ,\phi ) =
(-1)^{\hat{N}/2} (\hat{N} - 2 \hat{\mathbf{J}}
\cdot \mathbf{n}+ \hat{\openone}) \
\hat{U} (\theta ,\phi ) \
\exp(-i \pi \hat{J}_{z}) \
\hat{U}^{\dagger }(\theta ,\phi ) .
\end{equation}
This Wigner function will be examined for a variety
of states in the next section. However, it seems
pertinent to compare before these results with previous
approaches using the machinery of SU(2) quasidistributions.
One can immediately check that both give the same
$Q$ function, meanwhile the corresponding Wigner
functions are different. Once again, the symmetrical
order is extraordinarily sensitive to any fingerprint
of nonclassical behavior. Without going into mathematical
details, we merely quote that, as shown in Ref.~\cite{Kli02},
the SU(2) Wigner kernel can be represented as
\begin{equation}
\hat{w}_{\mathrm{SU(2)}}^{(0)}(\theta , \phi ) =
\hat{U}(\theta ,\phi ) \ F(\hat{J}_{z}) \
\hat{U}^\dagger(\theta ,\phi ),
\end{equation}
where $F(\hat{J}_{z})$ is an operator whose expression
is of little interest for our purposes here. It turns
out, however, that in the asymptotic case of large
dimensions of the representation ($J \gg 1$),
$F(\hat{J}_z)$ tends to the parity operator
on the sphere: $F(\hat{J}_{z}) \rightarrow
\ \exp (-i \pi \hat{J}_{z} )$. The SU(2) Wigner
function coincides thus with (\ref{wop}), except
for the factor the appears premultiplying, which
does not play any relevant role. So, in the classical
limit, both approaches give the same result.
Nevertheless, we mention that the SU(2) quasidistributions
have been not determined experimentally yet, at
difference of the easy measurability of the $s$-ordered
quasidistributions for the single polarization modes.

We finally note that the operator (\ref{wop}), when
restricted to a single SU(2) invariant subspace, does not
contain complete information: in other words, it cannot
be inverted to obtain the density matrix.

\section{Examples and concluding remarks}

To gain further insights into this formalism, we
shall particularize the Wigner function (\ref{wop})
for several states of interest. For simplicity,
we assume a pure state $| \Psi \rangle$, so
that the Wigner function is simply
\begin{equation}
W (\theta , \phi )= \Tr [
\hat{\varrho} (\theta ,\phi ) \
(-1)^{\hat{a}_V^\dagger  \hat{a}_V}
( 2 \hat{a}_H^\dagger \hat{a}_H +
\hat{\openone})],
\end{equation}
where
\begin{equation}
\hat{\varrho}(\theta ,\phi ) =
\hat{U}^\dagger (\theta ,\phi ) \
|\Psi \rangle \langle \Psi | \
\hat{U}(\theta ,\phi ).
\end{equation}

We recall that, for fixed $N$, the states $| N, k \rangle$
$(k= 0, \ldots, N$) span a $(N+1)$-dimensional invariant
subspace, wherein the action of the operators (\ref{gensu2})
is standard. In consequence, we expand $| \Psi \rangle$
in this basis
\begin{equation}
\label{decinvsub}
| \Psi \rangle =
\sum_{N=0}^\infty \sum_{k=0}^N
\Psi_{Nk} | N, k \rangle ,
\end{equation}
so that the action of $\hat{U}(\theta, \phi)$
on $| \Psi \rangle $ can be easily calculated.
The resulting Wigner function takes the form
\begin{eqnarray}
\label{w2}
W (\theta ,\phi ) & = & \sum_{N=0}^\infty
\sum_{k,m,n=0}^N (-1)^n \,
(2N -2n + 1)
\nonumber \\
& \times &
\Psi_{Nk} \Psi_{Nm}^\ast \, e^{i\phi (k-m)}
\, d_{nk}^N (- \theta ) d_{mn}^N (\theta ),
\end{eqnarray}
where $d_{nk}^N (\theta )$ is the Wigner
$d$ function
\begin{equation}
d_{k^\prime k}^N (\theta )  = \langle N, k^\prime |
\exp ( i \theta \hat{J}_y ) |N, k \rangle ,
\end{equation}
whose properties have been extensively studied~\cite{Var88}.
Using such properties and after some lengthy, but
otherwise straightforward calculations,
we finally get
\begin{eqnarray}
\label{wfin}
W(\theta , \phi ) & = & \sum_{N=0}^\infty
\sum_{m, n=0}^N (-1)^{n}
[ N + 1 + (m + n -N)/\cos \theta ]
\nonumber \\
& \times & \Psi_{Nm} \Psi_{Nn}^\ast
\, e^{i\phi (m-n)} \, d_{nm}^{N}(-2\theta ).
\end{eqnarray}
It can be  checked that $W(\theta , \phi )$
is properly normalized
\begin{equation}
\frac{1}{4\pi } \int d\Omega \
W(\theta ,\phi ) = 1.
\end{equation}

We first consider the  case of  (quadrature) coherent
states
\begin{equation}
| \Psi \rangle = | \alpha_H \rangle _H
\otimes | \alpha_V \rangle_V .
\end{equation}
We take both oscillators with the same
amplitude and relative phase $\varphi$; i.e.,
$\alpha_H = r/\sqrt{2}$ and $\alpha_V = r/\sqrt{2}
e^{-i \varphi}$. This is perhaps the most
interesting situation as polarization is concerned.
The decomposition in invariant subspaces reads
\begin{equation}
\Psi_{Nk} = e^{- |\alpha |^2}
\frac{|\alpha |^N}{\sqrt{( N - k)! k!}}
e^{i k \varphi},
\end{equation}
Note that this state is separable and $\Psi_{Nk}$ is
just the product of the photon-number amplitudes in
each polarization mode. The sums in (\ref{wfin}) can
be carried out explicitly, with the result
\begin{eqnarray}
\label{Wiguff}
W(\theta , \phi ) = \left \{ r^2
[ 1 + \sin \theta \cos ( \phi + \varphi ) ] + 1 \right \}
\exp \left \{ -2 r^2 [ 1- \sin \theta
\cos ( \phi + \varphi )] \right \} . \nonumber \\
\end{eqnarray}

To examine how the quantum character of the state
reflects itself in the properties of the Wigner
function, we plot the distribution~\cite{Dow94}
\begin{equation}
\label{funit}
f(\theta, \phi) = 1 + \frac{W (\theta, \phi)}
{ \langle \hat{N} \rangle }
\end{equation}
on the unit sphere. Notice that, with our parametrization,
if the unit vector $\mathbf{n}$ points in the positive direction
$x$, $y$, or $z$ the state is right-circularly, linearly at 45$^\circ$,
or horizontally polarized light, respectively. In Fig.~1 we have plotted
the normalized probability distribution $f(\theta, \phi)$
for the coherent case, with $r=5$ and the relative phase
$\varphi = \pi/2$, which classically corresponds to circularly
polarized light. What we see in the figure is indeed a Gaussian
distribution centered at the corresponding classical
point.

\begin{figure}
\centering
\resizebox{0.60\columnwidth}{!}{\includegraphics{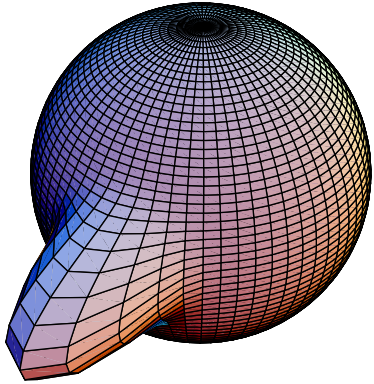}}
\caption{Spherical plot of the normalized distribution~(\ref{funit})
when the two polarization modes are in a coherent state.
We take both amplitudes equal to $r=5$ and the relative phase
$\varphi=\pi/2$. The Gaussian is centered at the classical
point corresponding to right-circularly polarized light.}
\end{figure}

Next we consider coherent squeezed states in both
polarization modes. In consequence, we have
\begin{equation}
| \Psi \rangle = | \alpha_H, \xi_H \rangle_H
\otimes | \alpha_V, \xi_V \rangle_V \, ,
\end{equation}
where
\begin{equation}
| \alpha, \xi \rangle = \hat{S} (\xi ) | \alpha \rangle .
\end{equation}
Here $| \alpha \rangle$ denotes a coherent state and
\begin{equation}
\hat{S}(\xi) =
\exp \left[ - \frac{1}{2}
( \xi^\ast \hat{a}^2 -
\xi \hat{a}^{\dagger 2} ) \right ]
\end{equation}
is the squeeze operator. The complex number
$\xi$ is known as the squeezing parameter and is
usually expressed $\xi = | \xi | e^{i \vartheta}$: while
$ | \xi |$ measures the squeezing of the fluctuations,
$\vartheta$ measures the direction in which such a
squeezing takes place.

The state is again separable and the decomposition
(\ref{decinvsub}) is now
\begin{equation}
\Psi_{Nk} = {}_H\langle N - k | \alpha_H, \xi_H \rangle_H
\; {}_V\langle k | \alpha_V, \xi_V \rangle_{V} .
\end{equation}
The photon-number amplitude for each polarization
mode is
\begin{equation}
\langle k | \alpha, \xi \rangle  =
\frac{[\nu / (2 \mu )]^{k/2}}
{\sqrt{\mu \,  k!}}
\exp \left [ - \frac{1}{2}
\left ( |\alpha |^2 - \frac{\nu^\ast}{\mu}
\alpha^2 \right) \right ]
\mathrm{H}_{k} \left( \frac{\alpha }{\sqrt{2\mu \nu }}\right) ,
\end{equation}
where $\mathrm{H}_{k} (x)$ are the Hermite polynomials
and we have used the  notation
\begin{equation}
\label{munu}
\mu = \cosh | \xi | ,
\qquad
\qquad
\nu = e^{i \vartheta} \sinh | \xi | .
\end{equation}
In this case, we have found no simple closed expression
for the Wigner function (\ref{wfin}). Nevertheless,
numerical calculations are simple. In Fig.~2 we have
represented the normalized distribution (\ref{funit})
when both modes are in the same squeezed state with a
real coherent amplitude  $\alpha = 5$ and a squeezing factor
$|\xi | = 0.3$ in the direction $\vartheta = 0$. Apart from
a small deformation of the sphere, we see two symmetrical
peaks with elliptical contours that indeed represent
squeezing of the fluctuations.

\begin{figure}
\centering
\resizebox{0.60\columnwidth}{!}{\includegraphics{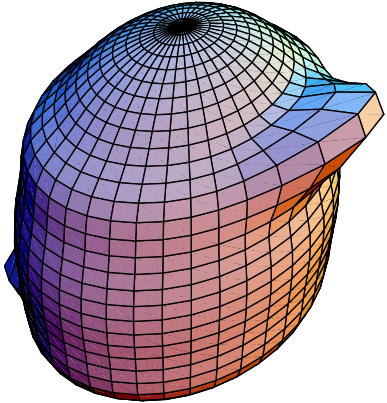}}
\caption{Spherical plot of the normalized distribution~(\ref{funit})
when the two polarization modes are in a squeezed state.
We take both modes in the same state, with a coherent amplitude
$\alpha = 5$  and a squeezing factor $| \xi | = 0.3$ in the direction
$\vartheta = 0$.}
\end{figure}

To give another simple but illustrative example,
we consider a two-mode squeezed vacuum state
\begin{equation}
|\Psi \rangle = \hat{S} (\xi) (
|0 \rangle_H \otimes |0 \rangle_V ) ,
\end{equation}
where the two-mode squeeze operator is
\begin{equation}
\hat{S} (\xi) =
\exp \left ( \xi^\ast \hat{a}_H \hat{a}_V
- \xi \hat{a}_V^\dagger \hat{a}_H^\dagger
\right) \, ,
\end{equation}
and $\mu$ and $\nu$ are defined as in
equation (\ref{munu}). The decomposition
in invariant subspaces for this state is
\begin{equation}
\Psi_{Nk}= \delta_{N, 2k}
\frac{1}{\mu} \left ( - \frac{\nu}{\mu} \right )^k .
\end{equation}
The Wigner function (\ref{wfin}) can be expressed
again in a closed form:
\begin{equation}
\label{Wtwom}
W (\theta ,\phi ) = \frac{1}{|\mu |^2}
\frac{1 - | \nu/\mu |^2}{[1 + 2 |\nu / \mu|
\cos (2 \theta) + |\nu / \mu|^2 ]^{3/2}} .
\end{equation}

\begin{figure}
\centering
\resizebox{0.60\columnwidth}{!}{\includegraphics{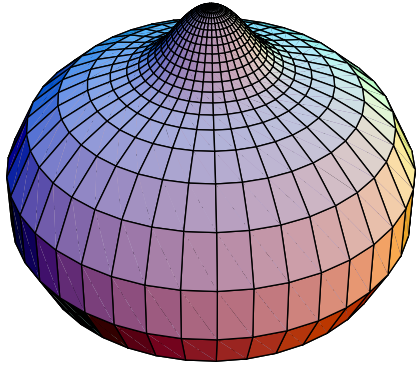}}
\caption{Spherical plot of the normalized distribution~(\ref{funit})
for a two-mode squeezed vacuum state with a squeezing parameter $| \xi | = 0.9$.}
\end{figure}

In Fig.~3 we have plotted the normalized
distribution corresponding to a two-mode squeezed
vacuum with $|\xi |  = 0.9$. We see the presence of two Gaussian
cups centered at the north and south poles and also
a  belt around the equator of the unit sphere.
This state can be seen as arising mainly from these
contributions and is rotationally symmetric, as
(\ref{Wtwom}) is independent of $\phi$.

Finally, we consider the propagation of light in a
Kerr medium. If initially both polarization modes
are in coherent states of amplitudes $\alpha_H$
and $\alpha_V$, the state at time $t$ can be
written as~\cite{Mil86,Kit86,Tan91,Tar93,Var93}
\begin{equation}
\label{cat}
|\Psi (\tau) \rangle = \hat{T} ( \tau )
( | \alpha_H \rangle_H \otimes
| \alpha_V \rangle_V ) ,
\end{equation}
where
\begin{equation}
\hat{T} ( \tau ) = \exp \left ( i
\tau \hat{a}_H^\dagger \hat{a}_H \
\hat{a}_V^\dagger\hat{a}_V \right ) ,
\end{equation}
and $\tau = \chi t$, $\chi$ being a real parameter
proportional to the third-order nonlinear
susceptibility of the medium. The state
(\ref{cat}) is entangled, and its
expansion (\ref{decinvsub}) can be
easily worked out
\begin{equation}
\label{PsiKerr}
\Psi_{Nk} = e^{i \tau (N - k) k} \
e^{-(|\alpha_H|^2 + |\alpha_V|^2)} \
\frac{| \alpha_H |^{N-k} | \alpha_V |^k} {\sqrt{(N - k)! k!}} .
\end{equation}

\begin{figure}
\centering
\resizebox{0.60\columnwidth}{!}{\includegraphics{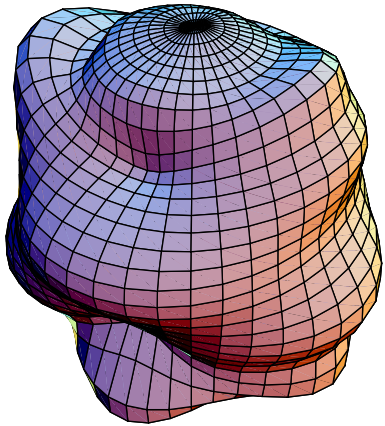}}
\caption{Spherical plot of the normalized distribution~(\ref{funit})
for light propagating in a Kerr medium. Initially the state is coherent
in both polarization modes with amplitude $r=5$ and the (adimensional)
propagation time is $\tau = \pi/2$.}
\end{figure}

The presence of the quadratic phase in (\ref{PsiKerr})
prevents again from obtaining an analytical form for
the Wigner function. In Fig.~4 we show the normalized
distribution on the sphere for a coherent amplitude
$ r = 5$ equal in both polarization modes and
$\tau = \pi/2$. The complicated phase dynamics
predicted by the theory manifests itself in
a complex pattern of well resolved peaks.

In summary, what we expect to have accomplished
in this paper is to present a simple alternative
phase-space formalism for polarization on the
Poincar\'e sphere, based on $s$-ordered
quasidistributions for the two basic polarization
modes. These results may have interesting experimental
consequences in order to implement a feasible
experimental procedure for determining polarization
properties.

\ack
The work of A. B. Klimov is partially supported
by the Grant PROMEP/103.5/04/1911.

\end{document}